\documentclass[pre,twocolumn,showpacs,groupedaddress,notitlepage,superscriptaddress,floatfix,nofootinbib,preprintnumbers]{revtex4-2}
\usepackage{subfigure}
\usepackage{braket}
\usepackage{tikz}
\usepackage{amsmath}
\usepackage{listings}
\usepackage[ruled]{algorithm} 
\usepackage{algpseudocode}
\usepackage{amsfonts}
\usepackage{float,mathdots}

\usepackage{multirow}
\usepackage{array}
\usepackage{amssymb}
\usepackage{accents}
\usepackage{amsthm}

\usepackage{hyperref}
\usepackage{blkarray}
\usepackage{svg}
\usepackage[dvipsnames]{xcolor}
\usepackage{booktabs}
\usepackage{makecell} 
\newcommand{\dd}{\mathrm{d}}

\usepackage{appendix}

\hypersetup{
    colorlinks=true,   
    linkcolor=cyan,    
    citecolor=magenta, 
    filecolor=magenta, 
    urlcolor=cyan,     
    runcolor=cyan
}
\usepackage[capitalise]{cleveref} 
\newcommand{\appropto}{\mathrel{\vcenter{
  \offinterlineskip\halign{\hfil$##$\cr
    \propto\cr\noalign{\kern2pt}\sim\cr\noalign{\kern-2pt}}}}}

\begin{document}

\title{Optimal heterogeneous diffusivity profiles for mean first-passage minimization}
\author{Luiz Menon}
\affiliation{Statistical and Interdisciplinary Physics Division, Centro Atómico Bariloche (CNEA), R8402AGP Bariloche, Argentina}
\email{luiz.menon@ib.edu.ar}

\begin{abstract}
We investigate how spatially varying diffusivity can be
optimized to minimize the mean first-passage time for escape
from a spherical shell with an inner reflecting boundary
and an outer absorbing boundary. Under a fixed
volume-averaged diffusivity, we derive a variational condition
for arbitrary spatial dimension and stochastic integration prescription.
For particles starting at the inner boundary, we obtain
exact optimal profiles in one dimension, with homogeneous
diffusion being optimal under the Stratonovich prescription.
In higher dimensions, the optimal Stratonovich profile
becomes decreasing. The optimal It\^o profile is decreasing
in all dimensions, whereas the optimal H\"anggi--Klimontovich
profile is increasing in one and two dimensions and can
exhibit a maximum in dimensions greater than two.
For particles starting beyond the inner boundary, optimal
profiles can exhibit a discontinuity at the initial
position, with constant diffusivity in the inner region.
Numerical optimization complements these results, showing
how geometry, stochastic prescription, and initial position
jointly determine the diffusivity distribution that
minimizes the mean escape time.
\end{abstract}
\maketitle

\section{Introduction}
Modifying how particles explore their surroundings provides a way
to accelerate escape from confined regions or encounters with
targets~\cite{benichou2011}. Both problems can be characterized by
the mean first-passage time (MFPT), defined as the average time
required for a stochastic particle to leave a region or reach
a target for the first time~\cite{redner2001guide,palyulin2014PNAS}.
Target-search processes arise in contexts ranging from proteins
locating specific sites on DNA~\cite{Mirny_2009,Chen_2019,Bhattacherjee2014}
to living organisms searching for
resources~\cite{BartumeusEcology2005,ricardoPRL,Kazimierski2016}. Strategies such as Lévy walks~\cite{viswanathan1999optimizing}, intermittent search~\cite{benichou2011}, and stochastic resetting~\cite{majundarPRL2011} have been explored to improve search efficiency under different conditions.

In confined systems, diffusive transport can limit the kinetics of chemical reactions by determining the time required for reactants to encounter one another~\cite{redner2001guide}. Consequently, it is important to understand how the geometry and properties of the medium influence first-passage times~\cite{benichou2010geometry}. In this context, heterogeneous diffusion~\cite{BarkaiPRE2019,Sandev_2022,PozoPRL2024}, characterized by spatial variations in diffusivity, can alter the time required for a particle to reach a target. Eukaryotic cells provide a paradigmatic example of heterogeneous media, where crowding and cytoskeletal organization can affect the MFPT and, consequently, the diffusive transport of intracellular cargo~\cite{Kuhn2011,CherstvySoft2014,ANDO20151574, Lanoiselee2018,HAFNER20181420,Xiang2020}. Describing such dynamics within the Langevin framework requires specifying the stochastic prescription (or interpretation), since spatially dependent diffusivity gives rise to multiplicative noise. Different interpretations can lead to distinct dynamics and, as a result, different mean first-passage times.

The use of spatially varying diffusivity coefficient to reduce the MFPT is discussed in Ref.~\cite{GodecPRE2015}, which considers a spherically symmetric system composed of two concentric regions with different diffusivities, featuring an absorbing inner boundary and a reflecting outer boundary. Using the Hänggi–Klimontovich prescription, it was shown that, under the constraint of a fixed volume-averaged diffusion coefficient, an appropriate ratio between the diffusivities of the two regions minimizes the MFPT.

In systems with this geometry, the MFPT and residence times in each region vary according to the adopted stochastic integration prescription, highlighting the relevance of this choice~\cite{vaccario2015first}. Related studies have also examined how spatial heterogeneity and stochastic prescriptions affect search efficiency in one-dimensional systems~\cite{santos_menon2022pre, menon2023bounded, menon2024PRE,LawleySearchPRE2026,LawleyPRL2026}. In this context, a question arises as to whether there is a heterogeneous diffusivity profile that makes the MFPT as short as possible and how this profile depends on the stochastic integration prescription and the spatial dimension.

In this work, we address this question for isotropic diffusion in a
$d$-dimensional spherical shell with an inner reflecting boundary
and an outer absorbing boundary. We consider a radial diffusivity
$D(r)$ with a fixed volume average and stochastic prescriptions
parameterized by $0 \leq \alpha \leq 1$. We formulate a variational
problem to minimize the MFPT to the absorbing boundary, focusing
first on particles starting at the inner boundary and then
considering other initial positions.

In one dimension, we obtain explicit expressions for the optimal
profile and the corresponding escape efficiency for arbitrary
$\alpha$. In higher dimensions, we combine analytical results
for selected prescriptions with numerical optimization for the
remaining ones. These results reveal how geometry and stochastic
interpretation shape the optimal diffusivity distribution,
including the emergence of nonmonotonic behavior.

The paper is organized as follows. In Sec.~\ref{sec:fps}, we introduce the stochastic model, the expression for the mean first-passage time, and the definition of the relative escape efficiency. In Sec.~\ref{sec:optimization}, we formulate the variational problem to determine the diffusivity profiles that minimize this time subject to a fixed volume-averaged diffusivity. We analyze the one-dimensional and higher-dimensional cases, examining the roles of the stochastic interpretation and the particle's initial position. Finally, Sec.~\ref{sec:conclusions} presents our concluding remarks.

\section{First-passage statistics}
\label{sec:fps}

We now address the escape problem between two spherical shells with heterogeneous diffusion.
Considering isotropic diffusion in $d$ spatial dimensions, governed by the following stochastic differential equation
\begin{equation}
    \frac{d\mathbf{x}(t)}{dt}
   =
    \sqrt{2D(\mathbf{x})}\,\boldsymbol{\xi}(t),
    \label{eq:lang1}
\end{equation}
where $\mathbf{x}(t)\in\mathbb{R}^d$ denotes the particle position and $D(\mathbf{x})$ is a position-dependent diffusion coefficient. The term $\boldsymbol{\xi}(t)$ is a $d$-dimensional Gaussian white noise with zero mean, $\langle\xi_i(t)\rangle=0$, and delta correlation $  \langle\xi_i(t)\xi_j(t')\rangle
    =\delta_{ij}\delta(t-t').$ 

Importantly, because the white noise is multiplicative, the stochastic integral in Eq.~\eqref{eq:lang1} is not uniquely defined. More precisely, it must be interpreted as
\begin{align}
   \int_0^t g(\mathbf{x}(t'))\,\boldsymbol{\xi}(t')dt' & = \int_0^t g(\mathbf{x}(t'))\,d\mathbf{W}_{t'}\nonumber\\
    &=
    \lim_{N\to\infty}
    \sum_{k=1}^{N}
    g(\mathbf{x}_k^{(\alpha)})
    \Delta\mathbf{W}_k,
\end{align}
where $g(\mathbf{x})=\sqrt{2D(\mathbf{x})}$, $\mathbf{W}_t$ is a \(d\)-dimensional Wiener process, $\Delta\mathbf{W}_k=\mathbf{W}_{t_k}-\mathbf{W}_{t_{k-1}}$, and the argument of $g$ is evaluated at
\begin{equation}
    \mathbf{x}_k^{(\alpha)}
    =(1-\alpha)\mathbf{x}(t_{k-1})
    +\alpha\mathbf{x}(t_k),
    \qquad 0\leq\alpha\leq1.
\end{equation}
Thus, $\alpha$ determines the relative weight assigned to the positions at the beginning and end of each time interval, and this parametrization is known as the generalized Stratonovich prescription or $\alpha$-prescription~\cite{hanggi1978stochastic,arenas2012hidden}. The choices $\alpha=0$, $\alpha=1/2$, and $\alpha=1$ correspond to the It\^o, Stratonovich, and Hänggi--Klimontovich prescriptions, respectively. For spatially uniform diffusivity, these prescriptions are equivalent, whereas for nonuniform $D(r)$ they generally lead to different stochastic dynamics.

The Fokker--Planck equation associated with the stochastic differential equation in Eq. (\ref{eq:lang1}) for the probability density $P(\mathbf{x},t)$ is given by
\begin{equation}
    \partial_t P(\mathbf{x},t)=\boldsymbol{\nabla}\cdot
    \left[D(\mathbf{x})^\alpha\boldsymbol{\nabla}
    \left(D(\mathbf{x})^{1-\alpha}P(\mathbf{x},t)\right)\right].
    \label{eq:fokker_planck}
\end{equation}

Assuming in addition radial symmetry, $D(\mathbf{x})=D(r)$, with $r=|\mathbf{x}|$, and a spherically symmetric probability density, the problem reduces to radial coordinates, yielding
\begin{equation}
    \partial_t P(r,t)=\frac{1}{r^{d-1}}\frac{\partial}{\partial r}
    \left[r^{d-1}D(r)^\alpha\frac{\partial}{\partial r}
    \left(D(r)^{1-\alpha}P(r,t)\right)\right].
    \label{eq:radial_fp}
\end{equation}
Here $P(r,t)$ is the density per unit $d$-dimensional volume; the radial density with respect to $\dd r$, since $\mathbf{x}=(r,\Omega)$, is $ p(r) = r^{d-1}\int d\Omega P(r,\Omega,t)=S_{d-1}r^{d-1}P(r,t)$, where $S_{d-1}=2\pi^{d/2}/\Gamma(d/2)$.

We consider the situation in which all particles are initially localized on the shell $r=r_0$, uniformly in angle, so that the initial condition is $P(r,0|r_0)=\delta(r-r_0)/(S_{d-1}r_0^{d-1})$ for $\bar r<r_0<R$. The domain is bounded by an inner reflecting boundary at $r=\bar r$ and an outer absorbing boundary at $r=R$, with $0<\bar r<R$. We assume $D$  is defined on $\bar r\leq r \leq R$ and $D(r)\ge 0$. The reflecting boundary imposes zero forward probability current,
\begin{equation}
    \left[-D(r)^\alpha\partial_r
    \left(D(r)^{1-\alpha}P(r,t|r_0)\right)\right]_{r=\bar r}=0,
\end{equation}
while $P(R,t|r_0)=0$.

We aim to compute the MFPT to the outer spherical shell. For this purpose, we apply the standard procedure based on the backward Fokker--Planck equation for the conditional probability density$P(r,t|r_0)$ \cite{Bray01062013}, namely  
\begin{equation}
    \partial_t P(r,t|r_0)=\frac{D(r_0)^{1-\alpha}}{r_0^{d-1}}
    \frac{\partial}{\partial r_0}\left[r_0^{d-1}D(r_0)^\alpha
    \frac{\partial P(r,t|r_0)}{\partial r_0}\right],
    \label{eq:backward_general}
\end{equation}
where $r_0$ denotes the initial radial position. 

Next, we introduce the survival probability, defined as the probability that a particle, initially located at $r_0$, remains within the interval $[\bar r,R]$ at time $t$
\begin{equation}
    Q(r_0,t)\equiv\int_{\bar r}^{R}p(r,t|r_0)\,\dd r.
    \label{eq:survival}
\end{equation}
Applying the definition of survival probability to the backward Fokker--Planck equation~\cite{redner2001guide}, Eq.~\eqref{eq:backward_general}, we obtain
\begin{equation}
    \partial_t Q(r_0,t)=\frac{D(r_0)^{1-\alpha}}{r_0^{d-1}}
    \frac{\partial}{\partial r_0}\left[r_0^{d-1}D(r_0)^\alpha
    \frac{\partial Q(r_0,t)}{\partial r_0}\right],
    \label{eq:backward_survival}
\end{equation}
where $Q(r_0,0)=1$ for $\bar r\leq r_0<R$, $\left.\partial_{r_0}Q(r_0,t)\right|_{r_0=\bar r}=0$, and $Q(R,t)=0$ for $t>0$.

The first-passage time density is defined as $\wp(t|r_0)=-\partial_t Q(r_0,t)$, and the MFPT is obtained via
\begin{equation}
    \begin{aligned}
    T(r_0)\equiv\int_0^\infty t\,\wp(t|r_0)\,\dd t=\int_0^\infty Q(r_0,t)\,\dd t,\qquad 
    \end{aligned}
    \label{eq:moments}
\end{equation}
From this, one derives differential equation
\begin{equation}
    \frac{D(r_0)^{1-\alpha}}{r_0^{d-1}}\frac{\dd}{\dd r_0}
    \left[r_0^{d-1}D(r_0)^\alpha\frac{\dd T}{\dd r_0}\right]
    =-1,
    \label{eq:moment_hierarchy}
\end{equation}
under the boundary conditions $T'(\bar r)=0$ and $T(R)=0$.
This equation can be integrated using the integrating factor
$r^{d-1}D(r)^\alpha$, yielding
\begin{equation}
    T(r_0)=\int_{r_0}^{R}\frac{y^{1-d}}{D(y)^\alpha}
    \left[\int_{\bar r}^{y}z^{d-1}D(z)^{\alpha-1}\,\dd z\right]\dd y.
    \label{eq:mfpt}
\end{equation}
Equivalently, reversing the order of integration gives
\begin{equation}
    T(r_0)=\int_{\bar r}^{R}z^{d-1}D(z)^{\alpha-1}
    \left[\int_{\max(r_0,z)}^{R}
    \frac{y^{1-d}}{D(y)^\alpha}\,\dd y\right]\dd z.
    \label{eq:mfpt2}
\end{equation}
We use these expressions as the starting point to optimize the diffusivity profile under a fixed volume-averaged diffusivity.

For a homogeneous diffusivity, $D(r)=D_0$, Eq.~\eqref{eq:mfpt} reduces to
\begin{equation}
    T_{\rm hom}=\frac{1}{D_0}
    \begin{cases}
        \displaystyle
            \frac{R^2-r_0^2}{4}
            -\frac{\bar r^{\,2}}{2}\ln\!\left(\frac{R}{r_0}\right),
        & d=2, \\[1.2em]
        \displaystyle
            \frac{R^2-r_0^2}{2d}
            -\frac{\bar r^{\,d}}{d(2-d)}
            \left(R^{2-d}-r_0^{2-d}\right),
        & d\neq 2.
    \end{cases}
    \label{eq:mfpt_constant_D}
\end{equation}
To quantify the effect of spatial heterogeneity, we define the relative escape efficiency with respect to the homogeneous baseline as
\begin{equation}
    \eta = 1 - \frac{T_{\rm het}}{T_{\rm hom}},
    \label{eq:efficiency}
\end{equation}
where $T_{\rm het}$ is given by Eq.~\eqref{eq:mfpt}, with both MFPTs evaluated for the same geometry and initial position. Positive values of $\eta$ indicate a reduction in the MFPT, negative values indicate an increase, and $\eta=0$ corresponds to equal MFPTs. In the constrained optimization introduced below, the homogeneous profile remains admissible, ensuring that the optimal escape efficiency satisfies $\eta_{\rm opt}\geq0$. Equality holds only when no admissible profile improves upon the homogeneous case.

\section{Constrained optimization of the diffusivity}
\label{sec:optimization}
We now ask how diffusivity should be distributed in space to minimize the mean first-passage time to the absorbing boundary. We take the particle to start at the reflecting boundary, $r_0=\bar r$. For any fixed positive diffusivity profile, this starting position maximizes the MFPT. Thus, minimizing $T(\bar r)$ is equivalent to minimizing the largest MFPT over all initial radial positions. We keep the geometry and the stochastic prescription $\alpha$ fixed while varying $D(r)$. To ensure a fair comparison with the homogeneous system, we require the volume-averaged diffusivity to remain equal to $D_0$, namely
\begin{equation}
    \langle D\rangle_V
    \equiv
    \frac{d}{R^d-\bar r^d}
    \int_{\bar r}^{R}
    r^{d-1}D(r)\,\dd r = D_0.
    \label{eq:avarege_constrain}
\end{equation}
The weight $r^{d-1}$ accounts for the increase in spherical shell volume with radius when $d>1$. Introducing a Lagrange multiplier $\lambda$ to enforce this condition, we define
\begin{equation}
    \mathcal{S}[D,\lambda]
    =
    T[D;\bar r]
    +
    \lambda
    \left[
        \langle D\rangle_V-D_0
    \right],
    \label{eq:lagrangian_general}
\end{equation}
where \(\lambda\) is a spatially uniform Lagrange multiplier. 

To formulate the optimization problem locally, we introduce
\begin{equation}
    I_d(r)
    =
    \int_{\bar r}^{r}
    z^{d-1}D(z)^{\alpha-1}\,\dd z,
    \label{eq:inner_integral_general}
\end{equation}
which satisfies
\begin{equation}
    I_d'(r)
    =
    r^{d-1}D(r)^{\alpha-1},
    \qquad
    I_d(\bar r)=0.
    \label{eq:Id_constraint}
\end{equation}
The mean first-passage time can then be written as
\begin{equation}
    T[D;\bar r]
    =
    \int_{\bar r}^{R}
    r^{1-d}D(r)^{-\alpha}I_d(r)\,\dd r.
\end{equation}

Treating \(I_d(r)\) as an independent field, we enforce
Eq.~\eqref{eq:Id_constraint} through an auxiliary Lagrange multiplier \(q(r)\).
The constrained functional therefore takes the local form
\begin{equation}
    \mathcal S
    =
    \int_{\bar r}^{R}
    \mathcal L
    \bigl(D,I_d,I_d',q;r\bigr)\,\dd r,
    \label{eq:local_action_general}
\end{equation}
with
\begin{align}
\label{eq:local_lagrangian_general}
    \mathcal L
    &=
    r^{1-d}D^{-\alpha}I_d\\
    &+
    \frac{\lambda d}{R^d-\bar r^d}
    r^{d-1}(D-D_0)+
    q
    \left[
        r^{d-1}D^{\alpha-1}\nonumber -I_d'
    \right].  
\end{align}
Stationarity of $\mathcal S$ with respect to the independent
fields $D(r)$, $I_d(r)$, and $q(r)$ gives the corresponding
Euler--Lagrange equations. Variation with respect to $q$
recovers Eq.~\eqref{eq:Id_constraint}, whereas the
Euler--Lagrange equation for $I_d$ yields
\begin{equation}
    q'(r)=-r^{1-d}D(r)^{-\alpha}.
    \label{eq:q_evolution}
\end{equation}
While $I_d(\bar r)=0$ is fixed, $I_d(R)$ has no prescribed
value. The corresponding natural boundary condition is
therefore
\begin{equation}
    \left.\frac{\partial\mathcal L}{\partial I_d'}\right|_{r=R}
    =-q(R)=0.
\end{equation}
Hence,
\begin{equation}
    q(r)=\int_r^R y^{1-d}D(y)^{-\alpha}\,\dd y
    \equiv J_d(r).
    \label{eq:outer_integral_general}
\end{equation}
Thus, $q(r)$ coincides with the inner integral in
Eq.~\eqref{eq:mfpt2} for $r_0=\bar r$.

Finally, since \(\mathcal L\) contains no derivative of \(D\), stationarity
with respect to \(D(r)\) reduces to
\begin{equation}
    \frac{\partial\mathcal L}{\partial D}=0,
\end{equation}
we obtain the following variational condition
\begin{equation}
    \frac{\alpha r^{2-2d}I_d(r)}
         {D(r)^{\alpha+1}}
    +
    \frac{(1-\alpha)J_d(r)}
         {D(r)^{2-\alpha}}
    =
    \frac{\lambda d}{R^d-\bar r^d}.
    \label{eq:optimality_general}
\end{equation}
This stationarity condition can also be obtained directly
by taking the first variation of the MFPT with respect
to the diffusivity profile under the fixed-average constraint ($\delta T/\delta D(r)
+\lambda \delta \langle D\rangle_V/{\delta D(r)}=0$).

For $0<\alpha<1$, the function
$f(D(z),D(y))=D(z)^{\alpha-1}D(y)^{-\alpha}$
has a positive definite Hessian with respect to its two
arguments. Thus, the MFPT (Eq.~\eqref{eq:mfpt}), a positively weighted integral
of $f$, is convex in $D$. For $\alpha=0,1$, the MFPT reduces
to $\int_{\bar r}^{R} w(r)/D(r)\,\mathrm{d}r$, with
$w(r)\geq0$, and is also convex since $1/D$ is convex
for $D>0$. The positivity and fixed-average constraints
define a convex admissible set. Therefore, any admissible
positive profile satisfying Eq.~\eqref{eq:optimality_general}
is a global minimizer.

Differentiating Eq.~(\ref{eq:optimality_general}) with respect to $r$, we obtain
\begin{equation}
	\frac{D_{\rm opt}'}{D_{\rm opt}}
	=
	\frac{
		2\alpha-1-2\alpha(d-1)r^{-d}I_d D_{\rm opt}^{1-\alpha}
	}{
		\alpha(\alpha+1)r^{1-d}I_d D_{\rm opt}^{1-\alpha}
		+(1-\alpha)(2-\alpha)r^{d-1}J_d D_{\rm opt}^{\alpha}
	},
	\label{eq:slope}
\end{equation}
this equation allows us to study whether the optimal diffusivity profile is increasing, decreasing, or has an extremum that could indicate a non-monotonic profile of $D_{\rm opt}(r)$.

The results of the following sections will be obtained from the Eq.~\eqref{eq:optimality_general} and Eq.~\eqref{eq:slope}. First, we treat the one-dimensional case ($d=1$) with arbitrary prescription ($0\leq\alpha\leq1$). Then, we consider the case of higher dimensions, specifying the prescription in the It\^o ($\alpha=0$), Stratonovich ($\alpha=1/2$), and H\"anggi--Klimontovich ($\alpha=1$) conventions.

\subsection{One-dimensional optimal profiles for a general stochastic prescription}
\label{sec:one_dimension}
For $d=1$, let $L=R-\bar r$, with reflection and the initial position at $r=\bar r$, and absorption at $r=R$. 
Consequently, the equation reduces to
\begin{equation}
    \frac{\alpha I_{1}(r)}{D(r)^{\alpha+1}}
    +\frac{(1-\alpha)J_{1}(r)}{D(r)^{2-\alpha}}= \frac{\lambda}{L}
    \label{eq:optimality_1d}
\end{equation}
Similarly, in one dimension case, Eq.~(\ref{eq:slope}) reduces to
\begin{equation}
	\frac{D'_{\rm opt}}{D_{\rm opt}}
	=
	\frac{
		2\alpha-1
	}{
		\alpha(\alpha+1)I_1 \,D^{1-\alpha}_{\rm opt}
		+(1-\alpha)(2-\alpha)J_1 D^{\alpha}_{\rm opt}
	},
	\label{eq:slope_1d}
\end{equation}
The condition in Eq.~(\ref{eq:slope_1d}) indicates three distinct scenarios, summarized below, regarding the growth or decay of the optimal diffusivity profile 
\begin{equation}
    \begin{cases}
    D_{\rm opt}'(r)<0, & 0<\alpha<1/2,\\
    D_{\rm opt}'(r)=0, & \alpha=1/2,\\
    D_{\rm opt}'(r)>0, & 1/2<\alpha<1.
    \end{cases}
    \label{eq:monotonicity_1d}
\end{equation}
For $\alpha < 1/2$, the derivative of the optimal profile is negative,
making $D_{\rm opt}$ decreasing. Thus, the diffusivity is larger near the
initial condition and smaller near the absorbing barrier. For $\alpha > 1/2$,
the behavior is the opposite. When $\alpha = 1/2$, the derivative vanishes,
indicating a constant optimal diffusivity profile, a feature that was also
observed in Refs.~\cite{santos2020pre,santos_menon2022pre,menon2023bounded}.

The optimal diffusivity profile is obtained by solving the variational condition, Eq.~(\ref{eq:optimality_1d}), together with the fixed mean diffusivity constraint, Eq.~(\ref{eq:avarege_constrain}), yielding
\begin{equation}
D_{\rm opt}(r)
=
\frac{3D_0}{2+\alpha\beta}
\Bigg[
\alpha\beta
+
\sqrt{
\beta^2\frac{R-r}{L}
+
\alpha^2\frac{r-\bar r}{L}
}
\Bigg],
\label{eq:optimal_profile_1d}
\end{equation}
where $\beta = 1-\alpha$. The detailed derivation of Eq.~(\ref{eq:optimal_profile_1d}) is provided in Appendix~\ref{app:1d_optimal}.

For the optimal profile corresponding to an arbitrary prescription,
$0\leq\alpha\leq1$, the optimal MFPT is
\begin{equation}
T_{\rm opt}
=
\frac{2L^2(2-\alpha)(1+\alpha)}{9D_0}.
\label{eq:T1_close}
\end{equation}
from which we identify
\begin{equation}
    \frac{T_{\rm opt}}{T_{\rm hom}}=\frac{4\,}{9}(2-\alpha)(1+\alpha).
\end{equation}
This ratio attains its minimum value, $8/9$, at the endpoints $\alpha=0$ and $\alpha=1$, and its maximum value, unity, at $\alpha=1/2$. Consequently, for any heterogeneous profile satisfying the same constraints used in the optimization,
\begin{equation}
T_{\rm het}\geq T_{\rm opt}\geq \frac{8}{9}T_{\rm hom}.
\label{eq:ratio_1d}
\end{equation}
Thus, within this admissible class, heterogeneity can reduce the MFPT by at most $1/9$  relative to the homogeneous case.

The relative optimal escape efficiency, defined in Eq.~(\ref{eq:efficiency}),
is
\begin{equation}
    \eta_{\rm opt}=\frac{1}{9}(2\alpha-1)^2.
    \label{eq:efficiency_1d}
\end{equation}
It reaches its maximum value, $1/9$, for the It\^o and
H\"anggi--Klimontovich prescriptions and vanishes for the
Stratonovich prescription where the optimal profile is the homogeneous one.

The invariance under $\alpha\leftrightarrow1-\alpha$ is
consistent with the conjugation of stochastic prescriptions
under time reversal in multiplicative-noise dynamics
\cite{arenas2012hidden,moreno2020state}.
In the present one-dimensional setting, this symmetry takes
the form of an exact MFPT identity: for a particle starting
at the reflecting endpoint $\bar r$, with absorption at $R$,
\begin{equation}
    T_\alpha[D]
    =
    T_{1-\alpha}[\widetilde D],
    \qquad
    \widetilde D(r)=D(R+\bar r-r).
    \label{eq:simmetry_alpha}
\end{equation}
Thus, reflecting the diffusivity profile and simultaneously
replacing $\alpha$ by $1-\alpha$ leaves the MFPT unchanged
\cite{menon2023bounded, menon2024PRE}.
This identity holds for arbitrary admissible diffusivity
profiles, not only for the optimal ones.
Since spatial reflection preserves the mean diffusivity
constraint, it also explains the symmetry of $T_{\rm opt}$
and $\eta_{\rm opt}$ about $\alpha=1/2$.
\begin{figure}
    \centering
\includegraphics[width=0.49\linewidth]{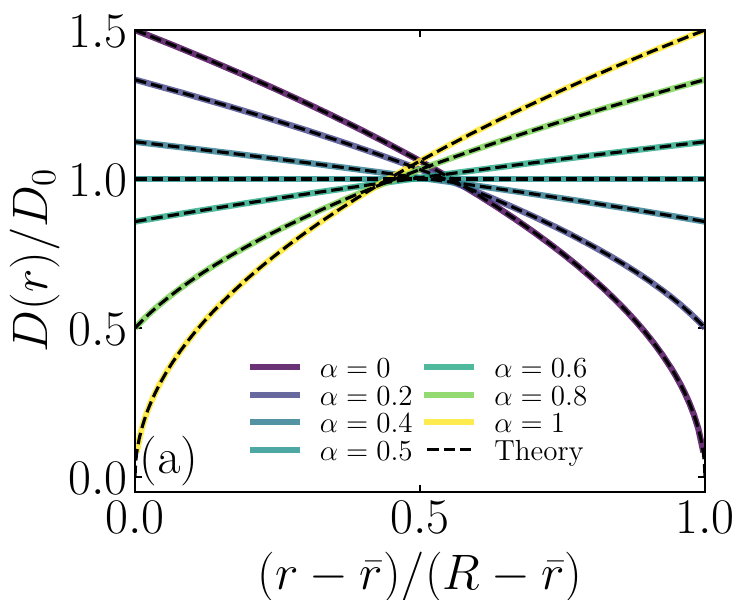}
\includegraphics[width=0.49\linewidth]{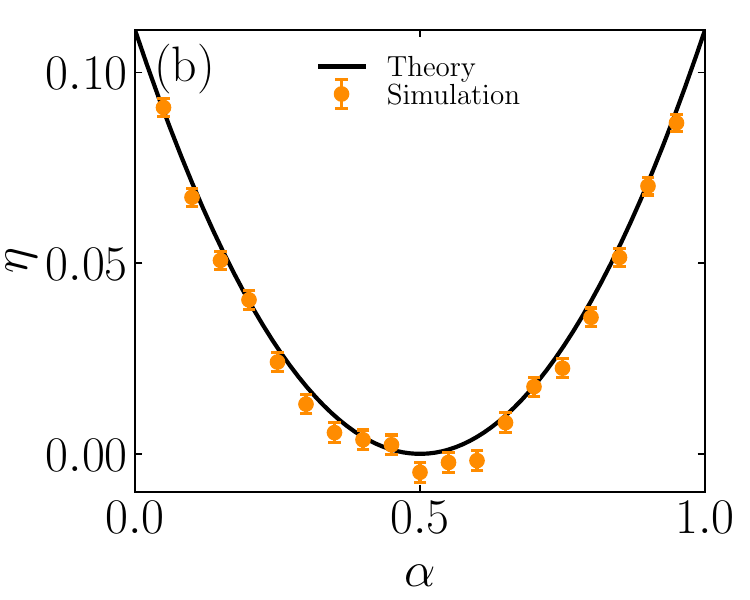}
\caption{
(a) Optimal one-dimensional diffusivity profiles minimizing the MFPT
from the reflecting boundary at $r=\bar r$ to the absorbing boundary
at $r=R$, at fixed mean diffusivity $D_0$.
Colored solid curves show the numerical optimization results, and
black dashed curves show the analytical predictions from
Eq.~\eqref{eq:optimal_profile_1d}
(b) Optimal relative escape efficiency. The black dashed line shows the
analytical prediction from Eq.~\eqref{eq:efficiency_1d}, and symbols
show Langevin simulation results obtained from $10^5$ independent
trajectories for each $\alpha$, details in Appendix \ref{app:langevin}.
Error bars represent one standard error of the mean.
}
    \label{fig:profiles_1d}
\end{figure}

Figure~\ref{fig:profiles_1d}(a) shows the optimal diffusivity
profiles obtained using the constrained gradient descent
method described in Appendix~\ref{ap:gradient}, together
with the analytical prediction from
Eq.~\eqref{eq:optimal_profile_1d}. The agreement between
the results for the values of $\alpha$ considered
confirms the consistency between the variational solution
and the numerical minimization of the MFPT.

Figure~\ref{fig:profiles_1d}(b) compares the optimal
relative escape efficiency predicted by
Eq.~\eqref{eq:efficiency_1d} with estimates obtained
from Langevin trajectory simulations using the optimal
profiles. Overall, the simulation results are in good
agreement with the analytical prediction.

\subsection{$d$-dimensional optimal profiles for select stochastic prescriptions}
\label{sec:d-dimensional}
In this section, we explore the generalized $d$-dimensional case for specific prescriptions. First, we consider the extreme cases $\alpha=0,1$, corresponding to the It\^o and H\"anggi--Klimontovich prescriptions, and finally the Stratonovich case.

\subsubsection{It\^o ($\alpha=0$)}
To determine the optimal diffusivity profile in the $d$-dimensional case under the It\^o prescription, we use the variation condition Eq.~(\ref{eq:optimality_general}) with $\alpha=0$, which straightforwardly gives
\begin{equation}
D_{\rm opt}(r)=D_0
\begin{cases}
\displaystyle
\frac{2(R^2-\bar r^2)}{\mathcal{A}_2}\sqrt{
\ln\left(\frac{R}{r}\right)},
& d=2, \\[10pt]
\displaystyle
\frac{(d-2)(R^d-\bar r^d)}{\mathcal{A}_d}\sqrt{r^{2-d}-R^{2-d}},
& d>2.
\end{cases}
\label{eq:optimality_d_ito}
\end{equation}
where $\mathcal{A}_2 =
\sqrt{\pi/2}R^2\operatorname{erf}(\sqrt{2\ln(R/\bar r)})
-2\bar r^2\sqrt{\ln(R/\bar r)}$,
and $\mathcal{A}_d =
dR^{d/2+1}B_z(3/2,(d+2)/(2(d-2)))$,
with $z=1-(\bar r/R)^{d-2}$.
Here, $B_z(a,b)$ denotes the incomplete beta
function \cite{AbramowitzStegun64}.

It is also possible to study the growth of the profile. By differentiating Eq.~(\ref{eq:optimality_d_ito}) or simply by applying $\alpha=0$ to Eq.~(\ref{eq:slope}), we obtain
\begin{equation}
 \frac{D_{\rm opt}'}{D_{\rm opt}}=-\frac{r^{1-d}}{2J_d(r)},
 \label{eq:ito_slope}
\end{equation}
since $J_d$ and $r^{1-d}$ are always positive, and the negative sign on the right-hand side of the equation ensures that the derivative is always negative. Therefore, the optimal profile for the It\^o prescription is always monotone and decreasing for any value of $d$.

\subsubsection{H\"anggi--Klimontovich ($\alpha=1$)}
For the H\"anggi--Klimontovich prescription, the variation condition also allows us to find the optimal profile, which is
\begin{equation}
    D_{\rm opt}(r) = \frac{3D_0}{2F_d}\left(\frac{r}{R}\right)^{1-d}\sqrt{\frac{r^d-\bar r^d}{R^d-\bar r^d}}
    \label{eq:optimality_d_hk}
\end{equation}
where $F_d = {}_2F_1\left(1-1/d,1;5/2;y\right)$
, with $y = 1- (\bar r/R)^d$, where ${}_2F_1$ denotes the Hypergeometric function \cite{AbramowitzStegun64}.
The derivative (Eq.~(\ref{eq:slope})) has the following form
\begin{equation}
 \frac{D'_{\rm opt}}{D_{\rm opt}
} =\frac{(2-d)r^d+2(d-1)\bar r^d}{2r(r^d-\bar r^d)}.
 \label{eq:slope_hk}.
\end{equation}

Since the denominator is positive for $\bar r<r<R$, the sign of the derivative is determined by the numerator.

For $d=2$, the numerator reduces to $2\bar r^2>0$. Consequently, the profile is strictly increasing inside the domain. For $d>2$, the condition $D_{\rm opt}'(r_c)=0$ gives
\begin{equation}
    r_c=\bar r
    \left[\frac{2(d-1)}{d-2}\right]^{1/d}.
    \label{eq:critical_radius}
\end{equation}
The derivative is positive for $r<r_c$ and negative for $r>r_c$. Thus, the profile has a single interior maximum only when $r_c<R$, or equivalently, when
\begin{equation}
    \frac{\bar r}{R}
    <
    \left[\frac{d-2}{2(d-1)}\right]^{1/d},
    \qquad d>2.
    \label{eq:hk_nonmonotonicity}
\end{equation}
If $R\leq r_c$, the profile remains strictly increasing inside the domain; in the case $R=r_c$, its derivative vanishes only at the outer boundary. Therefore, non-monotonicity depends on both the dimension and the ratio between the shell radii.

This behavior results from the competition between the factors $r^{1-d}$ and $\sqrt{r^d-\bar r^d}$ in Eq.~(\ref{eq:optimality_d_hk}): while the first decreases with radius for $d>1$, the second increases. Moreover, for fixed $\bar r$,
\begin{equation}
    \frac{r_c}{\bar r}
    =1+\frac{\ln 2}{d}+\mathcal{O}(d^{-2}),
    \qquad d \gg1.
\end{equation}
Therefore, in high dimensions, the maximum approaches the reflecting boundary. For any fixed ratio $R/\bar r>1$, this maximum lies inside the domain when $d$ is sufficiently large.
\subsubsection{Stratonovich ($\alpha=1/2$)}
For the intermediate value of $\alpha$, the variation condition, Eq.~(\ref{eq:optimality_general}), does not allow us to isolate the optimal profile, resulting in a transcendental integral equation. However, it is still possible to study the (de)growth by specializing Eq.~(\ref{eq:slope}) to $\alpha=1/2$, that is,
\begin{equation}
 \frac{D_{\rm opt}'}{D_{\rm opt}}
 =-\frac{4(d-1)}{3r}
 \frac{I_d(r)}{I_d(r)+r^{2d-2}J_d(r)}.
 \label{eq:slope_strat}
\end{equation}
Eq.~(\ref{eq:slope_strat}) also indicates that the optimal profile is decreasing for Stratonovich. This result contrasts with the one-dimensional case, in which the profile is constant.
\begin{figure}[h!]

    \centering
    \includegraphics[width=0.49\linewidth]{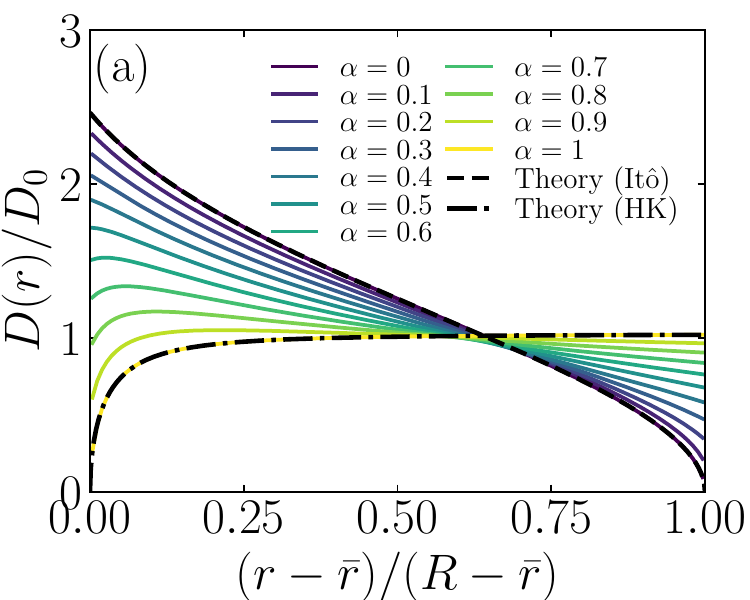}
       \includegraphics[width=0.49\linewidth]{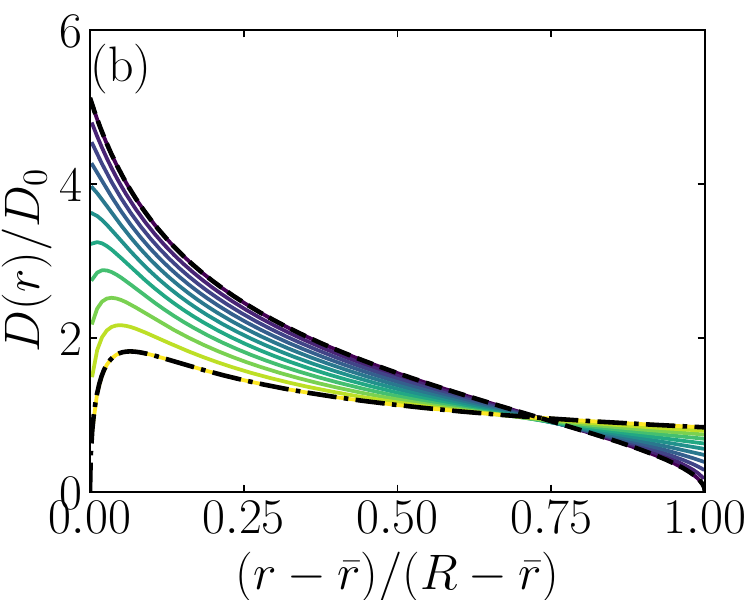}
\caption{
Optimal radial diffusivity profiles minimizing the mean first-passage
time from the reflecting boundary $r_0=\bar r=1$ to the absorbing
boundary $R=10$ in (a) $d=2$ and (b) $d=3$, at fixed volume-averaged
diffusivity $D_0$.
Colored solid curves: numerical optimization results for the $\alpha$ values indicated in panel (a).
Black dashed and dash-dotted curves depict analytical predictions for
It\^o ($\alpha=0$) and H\"anggi--Klimontovich ($\alpha=1$)
prescriptions, given by Eq.~\eqref{eq:optimality_d_ito}
and Eq.~\eqref{eq:optimality_d_hk}, respectively.
}
\label{fig:optimal_profiles}
\end{figure}

 Figures \ref{fig:optimal_profiles}(a) and \ref{fig:optimal_profiles}(b) show the numerically optimized diffusivity profiles for different values of $\alpha$ in two and three dimensions, respectively. The analytical predictions for the It\^o ($\alpha=0$) and H\"anggi--Klimontovich ($\alpha=1$) prescriptions are also shown and are in excellent agreement with the corresponding numerical results. For the Stratonovich prescription ($\alpha=1/2$), only numerical results are shown. 

For the two-dimensional case (Fig.~\ref{fig:optimal_profiles}(a)), the analytical results show that the optimal diffusivity profiles exhibit monotonic behavior for specific prescriptions. The optimal profile $D_{\rm opt}$ decreases for $\alpha=0$ and increases for $\alpha=1$. For the Stratonovich prescription ($\alpha=1/2$), although we do not have an analytical expression for the optimal profile $D_{\rm opt}$, the analysis of its monotonicity, provided by Eq.~\eqref{eq:slope_strat}, predicts a monotonically decreasing behavior, in agreement with the numerical results. However, while the optimal profiles are monotonic for $\alpha=0$, $1/2$, and $1$, the numerical results reveal the existence of an intermediate region, with $1/2<\alpha<1$, in which the optimal profile becomes nonmonotonic, exhibiting a maximum.

For $d>2$, as illustrated by the three-dimensional case in Fig.~\ref{fig:optimal_profiles}(b), the optimal profile for the H\"anggi--Klimontovich prescription is nonmonotonic, as theoretically predicted by Eq.~\eqref{eq:slope_hk} and Eq.~\eqref{eq:optimality_d_hk}. Furthermore, for this specific case ($d=3$), the numerical results reveal that the optimal profiles are also nonmonotonic for certain values of $\alpha$ in the interval $1/2<\alpha<1$.
\begin{figure}[h!]
    \centering
    \includegraphics[width=0.49\linewidth]{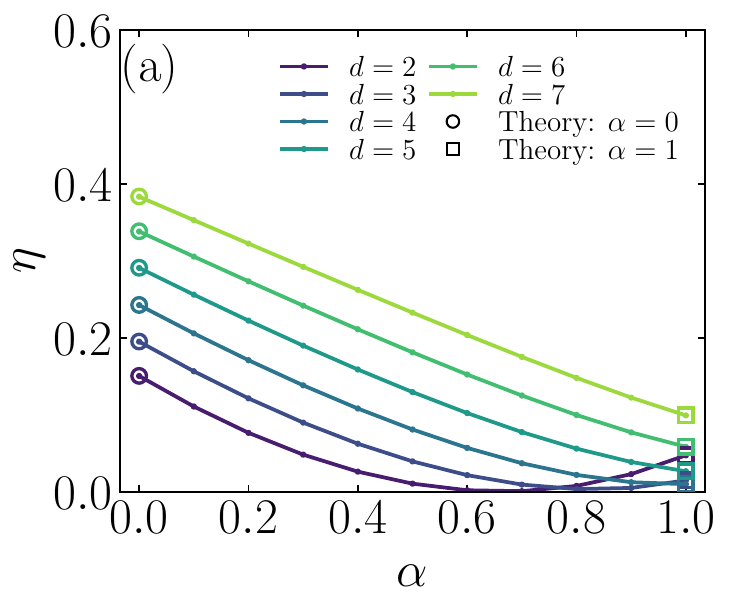}
       \includegraphics[width=0.49\linewidth]{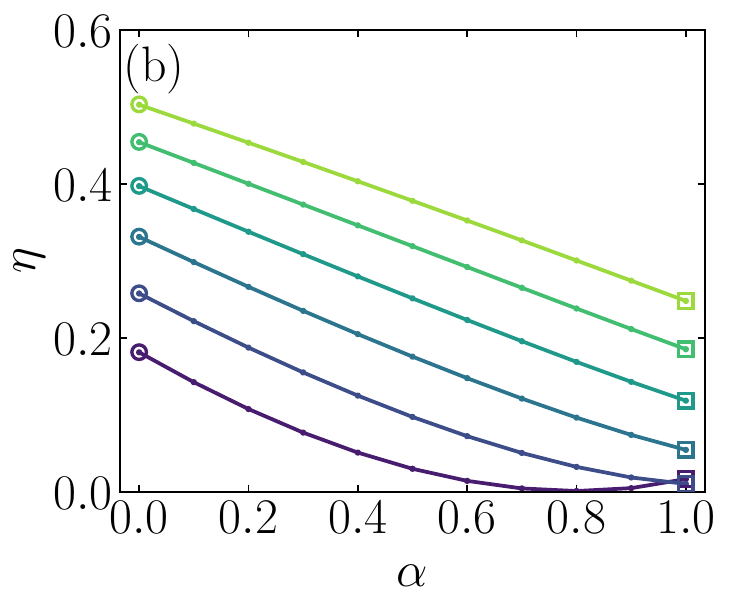}
\caption{
Optimal relative escape efficiency, Eq.~\eqref{eq:efficiency},
as a function of the stochastic prescription parameter $\alpha$
for different dimensions $d$.
The particle starts at the reflecting inner boundary
$r_0=\bar r=1$, with an absorbing outer boundary at
(a) $R=2$ and (b) $R=4$.
Colored curves show the numerical optimization results.
Open circles and squares indicate the analytical predictions
for the It\^o ($\alpha=0$) and H\"anggi--Klimontovich ($\alpha=1$)
prescriptions, respectively.
}\label{fig:eff_vs_alpha}
\end{figure}

The panels in Fig.~\ref{fig:eff_vs_alpha}(a) and Fig.~\ref{fig:eff_vs_alpha}(b) show the relative escape efficiency, defined by Eq.~\eqref{eq:efficiency}, as a function of the stochastic prescription—parameterized by $\alpha$—for different dimensions $d$ and external radii $R$.

In Fig. Fig.~\ref{fig:eff_vs_alpha}(a), for $R=2$, we observe that the highest relative escape efficiencies occur for values of $\alpha$ close to zero, regardless of the dimensionality. For higher dimensions ($d=4, 5, 6, 7$), the relative escape efficiency decreases monotonically as $\alpha$ increases. On the other hand, for $d=2$ and $d=3$, a minimum in the relative escape efficiency is observed at intermediate values of $\alpha$, where it becomes small but stays positive. This behavior is related to the fact that, in these cases, the optimal diffusivity profile approaches a constant profile, causing the optimized MFPT to approach that obtained for a homogeneous medium.

In Fig.~\ref{fig:eff_vs_alpha}(b), for $R=4$, the predominantly decreasing behavior of relative escape efficiency as a function of $\alpha$ is maintained for the dimensions analyzed, with the exception of $d=2$. In this case, a minimum in relative escape efficiency is observed once again, though it is shifted relative to the $\alpha$ value found for $R=2$.

Comparing panels (Fig.~\ref{fig:eff_vs_alpha}(a) and Fig.~\ref{fig:eff_vs_alpha}(b)), we observe that, for larger values of $d$, the relative escape efficiency tends to increase with the outer radius $R$. This indicates that optimal diffusivity profiles yield a greater relative reduction in the MFPT in larger domains, compared with the homogeneous case. Furthermore, for the parameters analyzed, the relative escape efficiency tends to increase with dimensionality. This behavior can be understood in terms of the increasing fraction of the system's volume concentrated near the outer boundary as $d$ increases. Under the constraint of fixed volume-averaged diffusivity, a small reduction in diffusivity in the outer region can therefore allow for a substantial increase in the inner region, which occupies a smaller fraction of the total volume. When this redistribution reduces the MFPT, it leads to a greater relative gain over the homogeneous case.
\subsection{Optimal diffusivity profiles for
\texorpdfstring{$r_0 \neq \bar{r}$}{r0 different from r-bar}}

For completeness, we consider an initial position
$\bar r<r_0<R$. Note that the integral \eqref{eq:mfpt} can be rewritten as
\begin{equation}
    T(r_0)=  \int_{\bar r}^{R}\frac{\Theta(y-r_0)y^{1-d}}{D(y)^\alpha}
    \left[\int_{\bar r}^{y}z^{d-1}D(z)^{\alpha-1}\,\dd z\right]\dd y.
    \label{eq:mfpt_heaviside}
\end{equation}
where $\Theta$ is the Heaviside step function.
Starting from Eq. \eqref{eq:mfpt_heaviside} and following the previous procedure, we find the Lagrangian
\begin{align}
    \mathcal L
    &=
    \Theta(r-r_0)\,r^{1-d}D^{-\alpha}I_d
    \nonumber\\
    &\quad+
    \frac{\lambda d}{R^d-\bar r^d}
    r^{d-1}(D-D_0)
    \nonumber\\
    &\quad+
    q\left[r^{d-1}D^{\alpha-1}-I_d'\right],
    \label{eq:local_lagrangian_r0}
\end{align}
Variation with respect to $q$ recovers
Eq.~\eqref{eq:Id_constraint}, whereas variation with
respect to $I_d$ gives
\begin{equation}
    q'(r)
    =
    -\Theta(r-r_0)\,r^{1-d}D(r)^{-\alpha}.
\end{equation}
Together with the natural boundary condition $q(R)=0$,
this yields
\begin{equation}
    q(r)
    =
    \int_{\max(r,r_0)}^{R}
    y^{1-d}D(y)^{-\alpha}\,\dd y
    \equiv J_{d,r_0}(r).
\end{equation}
Finally, stationarity with respect to $D(r)$ gives
\begin{equation}
    \frac{\alpha\Theta(r-r_0)r^{2-2d}I_d(r)}
         {D(r)^{\alpha+1}}
    +
    \frac{(1-\alpha)J_{d,r_0}(r)}
         {D(r)^{2-\alpha}}
    =
    \frac{\lambda d}{R^d-\bar r^d}.
    \label{eq:optimality_general_r0}
\end{equation}
For $r>r_0$, this condition coincides with
Eq.~\eqref{eq:optimality_general}. For $\bar r<r<r_0$,
it reduces to
\begin{equation}
    \frac{(1-\alpha)J_d(r_0)}
         {D(r)^{2-\alpha}}
    =
    \frac{\lambda d}{R^d-\bar r^d}.
\end{equation}
Consequently, for $0\leq\alpha<1$, any interior
optimal profile is constant below the initial position:
\begin{equation}
    D_{\rm opt}(r)=
    \left[
        \frac{(1-\alpha)J_d(r_0)
        \left(R^d-\bar r^d\right)}
        {\lambda d}
    \right]^{\frac{1}{2-\alpha}},
    \quad \bar r<r<r_0.
    \label{eq:D_optimal_in}
\end{equation}
Here, both $J_d(r_0)$ and $\lambda$ depend implicitly
on $\alpha$ and on the optimal profile.

The case $\alpha=1$ requires separate consideration.
In this case, the MFPT reduces to
\begin{equation}
    T(r_0)=
    \frac{1}{d}\int_{r_0}^{R}
    \frac{y^{1-d}\left(y^d-\bar r^d\right)}
         {D(y)}\,\dd y,
\end{equation}
and is independent of the diffusivity in the interior interval,
$\bar r\leq r<r_0$. Under the fixed-average constraint, the
diffusivity budget can therefore be concentrated in the exterior
interval, $r_0<r\leq R$, as the interior diffusivity approaches
zero. This allocation is consistent with transport favoring
the absorbing boundary at $R$, while suppressing excursions
into the low-diffusivity interior region, where motion would
be strongly slowed. However, the absence of such excursions
cannot be inferred from the MFPT alone.
\begin{figure}[h!]
    \centering
    \includegraphics[width=0.49\linewidth]{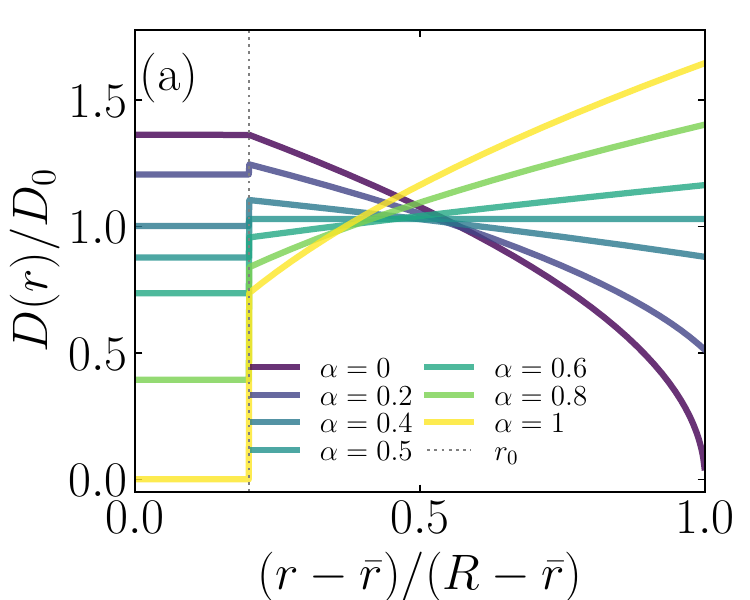}
       \includegraphics[width=0.49\linewidth]{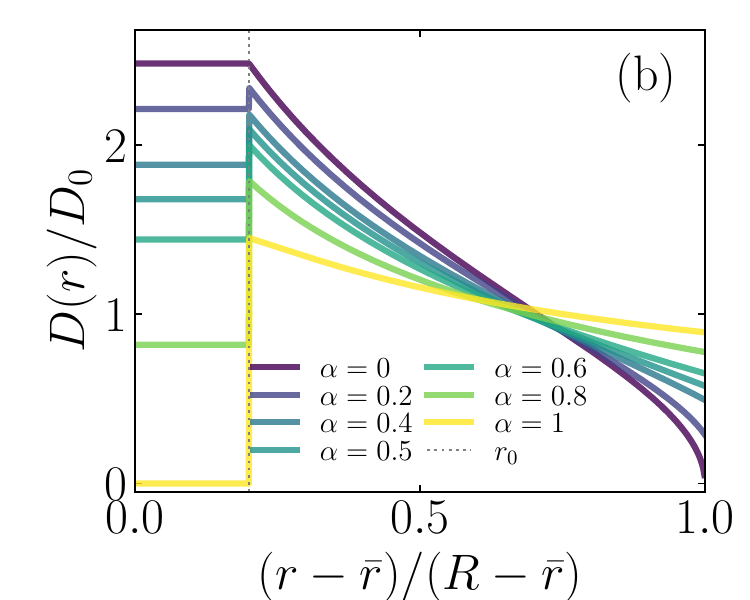}
\caption{Optimal radial diffusivity profiles minimizing the mean first-passage
time from the reflecting boundary at $  \bar r =1$ an initial condition $r_0=2$ to the absorbing
boundary $R=6$ in (a) $d=1$ and (b) $d=3$, at fixed volume-averaged
diffusivity $D_0$.}\label{fig:profiles_r0}
\end{figure}

Fig.~\ref{fig:profiles_r0}(a) shows the numerically optimized heterogeneous diffusivity profiles for $d=1$. The plateaus in the interior region ($\bar r\leq r<r_0$) are observed as predicted by Eq.~\eqref{eq:D_optimal_in}, for $0\leq\alpha<1$. Unlike the case where $r_0=\bar r$, for Stratonovich's prescription ($\alpha=1/2$), the optimal profile ceases to be homogeneous throughout the interval ($\bar r\leq r\leq R$) and becomes piecewise constant, with a discontinuity at $r=r_0$.
For the It\^o prescription ($\alpha=0$), the plateau corresponds to the maximum diffusivity over the entire interval, followed by a monotonically decreasing profile in the outer region ($r_0<r\leq R$). The profile is continuous, but its derivative is discontinuous at $r=r_0$. As the value of $\alpha$ increases, the level of the plateau in the inner region decreases. For prescriptions with $0\leq\alpha<1/2$, the diffusivity profile in the outer region is decreasing; the opposite occurs for prescriptions with $1/2<\alpha\leq1$. The constant value of diffusivity in the inner region approaches zero as $\alpha$ approaches unity, as discussed previously.

Fig.~\ref{fig:profiles_r0}(b) shows the numerically optimized profiles for the three-dimensional case. The discussion of the plateaus in the inner region for the one-dimensional case (Fig.~\ref{fig:profiles_r0}(a)) also is valid  here. However, unlike the one-dimensional case, the profiles in the three-dimensional case exhibit a monotonic decay as a function of the radius $r$ in the exterior region ($r_0<r\leq R$), for all the interpretations presented. In particular, for Stratonovich's prescription ($\alpha=1/2$), the profile in the outer region ceases to be constant and becomes decreasing, highlighting the effect of dimensionality on the optimum diffusivity. An inversion in the ordering of the profiles in this region is also observed: near $r_0$, the lowest values of $\alpha$ correspond to the highest diffusivities, while near the absorbing boundary $R$ the opposite occurs.

\begin{figure}[h!]
    \centering
    \includegraphics[width=0.49\linewidth]{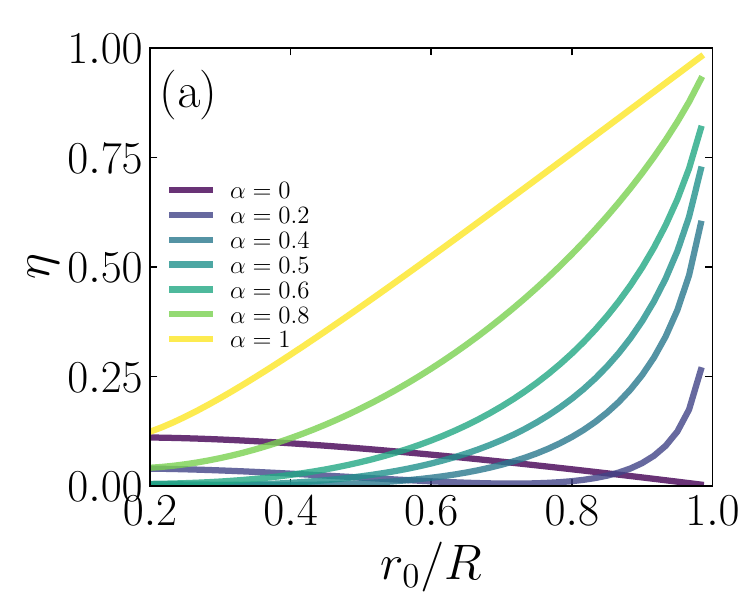}
       \includegraphics[width=0.49\linewidth]{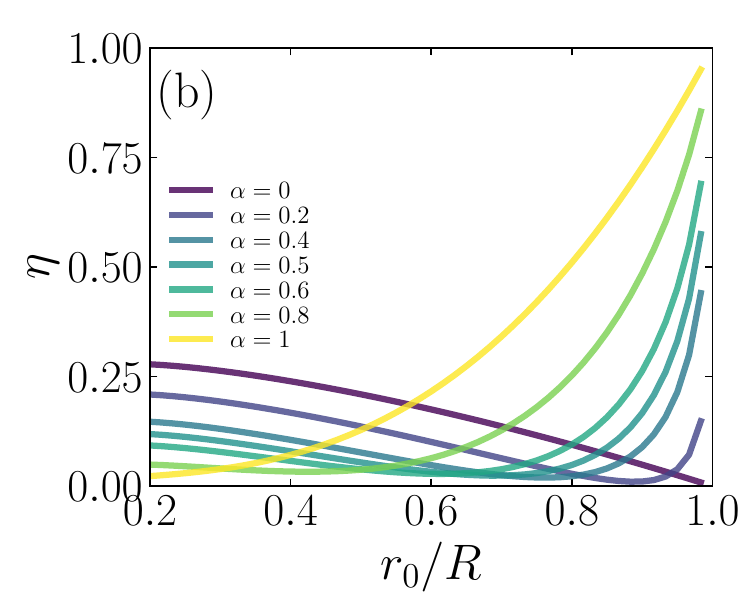}
\caption{Efficiency, $\eta(\alpha,r_0)$, as a function of the initial position $r_0$ for different values of $\alpha$, with $R=6$ and $\bar r=1$: (a) $d=1$ and (b) $d=3$. The diffusivity profile is numerically optimized independently for each pair $(\alpha,r_0)$.}\label{fig:eff_r_0}
\end{figure}
In Fig \ref{fig:eff_r_0}(a) the relative escape efficiency as function of initial radial position $r_0$ is shown for $d=1$. For the It\^o prescription ($\alpha=0$) the escape efficiency decreases  as the initial position approaches the absorbing boundary, indicating a reduction of advantage  of the optimized profile compared to the homogeneous one. For the H\"anggi-Klimontovich ($\alpha=1$), the escape efficiency increases and remains above the other curves for all interval of $r_0$. This behavior is associated with the possibility of concentrating all the amount of diffusivity in the other region in the interval $r_0\leq r \leq R$. For intermediate values of $\alpha$, a transition between these regimes is observed, with relative escape efficiency minima followed by growth close to the absorption boundary.  

In Fig.~\ref{fig:eff_r_0}(b), we show the relative escape efficiency
as a function of the initial position in three dimensions.
The ordering of the escape efficiencies reverses as $r_0$ varies.
Near the inner boundary, lower values of $\alpha$ provide
the greatest relative gains, whereas closer to the absorbing
boundary, higher values of $\alpha$ become more favorable.
For $0<\alpha<1$, the relative escape efficiency exhibits a minimum,
which shifts toward the outer boundary at $r=R$ as $\alpha$
decreases. These minimums identify initial positions for which diffusivity
optimization offers relatively little improvement over the
homogeneous medium. Under the fixed-average constraint,
reallocating diffusivity to the outer region ($r_0<r\leq R$)
requires reducing it in the inner region
($\bar r\leq r<r_0$). For $0<\alpha<1$, this reduction increases
the inner region's contribution to the MFPT, partially offsetting
the benefit of increased diffusivity in the outer region. 

Compared with the one-dimensional case (Fig.~\ref{fig:eff_r_0}(a)), the outer region
occupies a larger fraction of the total volume at the same
initial position. Consequently, concentrating the diffusivity
budget in this region produces a smaller increase in its
average diffusivity. This geometric effect helps explain
why the sharp increase in escape efficiency occurs closer to $R$
in higher dimensions.

\section{Final remarks}
\label{sec:conclusions}

To summarize, we studied the optimization of radial diffusivity profiles to minimize the MFPT in a domain bounded by an inner reflecting boundary and an outer absorbing boundary, subject to a fixed volume-averaged diffusivity.

A stationarity condition valid for any stochastic integration prescription parameterized by $0\leq\alpha\leq1$ and any spatial dimension was obtained through a variational formulation. We also defined a relative escape efficiency that quantifies the reduction in MFPT compared with a homogeneous medium.

For particles starting at the reflecting boundary, $r_0=\bar r$, the optimal one-dimensional diffusivity profile was found to decrease for $\alpha<1/2$, remain constant for $\alpha=1/2$ (the Stratonovich prescription), and increase for $\alpha>1/2$. We derived an expression for the optimal escape efficiency, which vanishes at $\alpha=1/2$, is symmetric about this point, and reaches its maximum at $\alpha=0$ and $\alpha=1$, where the relative reduction in MFPT is $1/9$.

For $d>1$, the optimal profile decreases under both the It\^o ($\alpha=0$) and Stratonovich ($\alpha=1/2$) prescriptions, unlike the constant profile obtained for Stratonovich in one dimension. Under the Hänggi--Klimontovich prescription ($\alpha=1$), the optimal profile increases for $d=2$, whereas for $d>2$ it can become nonmonotonic, depending on the ratio of the inner radius $\bar r$ to the outer radius $R$. Numerical optimization also reveals nonmonotonic optimal profiles for intermediate prescriptions with $1/2<\alpha<1$. The numerical profiles agree with the available analytical solutions.

Within the parameter ranges investigated, the escape efficiency gain tends to increase with both the spatial dimension and the ratio $R/\bar r$. These results show that the optimal diffusivity distribution depends jointly on the geometry, the stochastic prescription, and the constraint on the average diffusivity.

When the initial radius lies beyond the inner boundary, $r_0>\bar r$, the optimized profiles exhibit a plateau in the inner region ($\bar r\leq r<r_0$) for any stochastic prescription and spatial dimension. The plateau level decreases as $\alpha$ increases. Under the Hänggi--Klimontovich prescription ($\alpha=1$), the MFPT is independent of the diffusivity in this region, allowing the profile to vanish there. We also analyzed the escape efficiency gain as a function of the initial position, highlighting its dependence on the stochastic prescription and the spatial dimension.

Future work will explore the optimization of channel shapes within the Fick--Jacobs framework and optimal diffusivity profiles in wedge-shaped domains.

\begin{acknowledgments}
The author thanks Danilo Cius for valuable comments. This work was supported by a Simons Foundation Targeted Grant to Instituto Balseiro, Argentina.
\end{acknowledgments}

\section*{AI-assisted research statement}

AI tools were used to assist with grammar and language editing
and to help check mathematical derivations. The author reviewed
the AI-assisted content and takes full responsibility for the
accuracy and integrity of the manuscript.


\appendix
\section{One-dimensional optimal profile}
\label{app:1d_optimal}
We consider the one-dimensional case with the particle starting at the
reflecting boundary. We denote the spatial coordinate by $x$ and write
$L=R-\bar r$, $\beta=1-\alpha$, and $c=\lambda/L$.
We start from the Lagrangian in Eq.~\eqref{eq:local_lagrangian_general}
for $d=1$. Since it has no explicit dependence on $x$, we use the
Beltrami identity \cite{Arfken2005} and define the constant of integration
\begin{equation}
h
=
\mathcal{L} -I_1'
\frac{\partial \mathcal{L}}{\partial I_1'}
.
\end{equation}
Using $q=J_1$, we obtain
\begin{equation}
h
=
\frac{I_1}{D^\alpha}
+\frac{J_1}{D^\beta}
+c(D-D_0).
\label{eq:app_Hamiltonian}
\end{equation}
Another useful relation follows by multiplying the one-dimensional
variational condition, Eq.~\eqref{eq:optimality_1d}, by $D(x)$:
\begin{equation}
\alpha\frac{I_1}{D^\alpha}
+
\beta\frac{J_1}{D^\beta}
=cD.
\label{eq:app_stat_D}
\end{equation}
We now relate $c$ to the MFPT using the fixed mean diffusivity
constraint, Eq.~\eqref{eq:avarege_constrain}. Setting $d=1$ and
$r_0=\bar r$ in Eq.~\eqref{eq:mfpt} and interchanging the order
of integration, we obtain
\begin{equation}
T= \int_{\bar r}^{R}
\frac{I_1}{D^\alpha}\,\dd x
=
\int_{\bar r}^{R}
\frac{J_1}{D^\beta}\,\dd x.
\label{eq:mpft_order}
\end{equation}
Integrating Eq.~\eqref{eq:app_stat_D} and using
Eq.~\eqref{eq:avarege_constrain} yields
\begin{equation}
T=cD_0(R-\bar r).
\end{equation}
Integration of Eq.~\eqref{eq:app_Hamiltonian} then gives
\begin{equation}
h=2cD_0.
\label{eq:app_Hvalue}
\end{equation}
Consequently,
\begin{equation}
\frac{I_1}{D^\alpha}
+
\frac{J_1}{D^\beta}
=
c(3D_0-D).
\label{eq:app_sum}
\end{equation}

Differentiating Eq.~\eqref{eq:app_stat_D} and using
$I_1'=D^{-\beta}$ and $J_1'=-D^{-\alpha}$, we obtain
\begin{equation}
\left[
cD
+\alpha^2\frac{I_1}{D^\alpha}
+\beta^2\frac{J_1}{D^\beta}
\right]D'
=
\alpha-\beta.
\label{eq:app_before_elim}
\end{equation}

Since
\begin{equation}
\alpha^2=\alpha-\alpha\beta,
\qquad
\beta^2=\beta-\alpha\beta,
\end{equation}
the integral terms can be eliminated using
Eqs.~\eqref{eq:app_stat_D} and \eqref{eq:app_sum}. This gives
\begin{equation}
c\left[
(2+\alpha\beta)D
-3\alpha\beta D_0
\right]D'
=
\alpha-\beta.
\label{eq:app_closed}
\end{equation}

Equation~\eqref{eq:app_closed} implies
\begin{equation}
\frac{\dd}{\dd x}
\left[
D-
\frac{3\alpha\beta D_0}
{2+\alpha\beta}
\right]^2
=
\mathrm{const}.
\label{eq:app_affine}
\end{equation}
Thus, the squared shifted diffusivity varies linearly with $x$.

The endpoint values follow from
Eqs.~\eqref{eq:app_stat_D} and \eqref{eq:app_sum}. Since
\(I_1(\bar r)=0\),
\begin{equation}
D(\bar r)
=
\frac{3\beta}{1+\beta}D_0.
\label{eq:app_Dleft}
\end{equation}
Similarly, since \(J_1(R)=0\),
\begin{equation}
D(R)
=
\frac{3\alpha}{1+\alpha}D_0.
\label{eq:app_Dright}
\end{equation}

Using these endpoint values in Eq.~\eqref{eq:app_affine},
we find
\begin{equation}
\begin{split}
\left[
D(x)-
\frac{3\alpha\beta D_0}
{2+\alpha\beta}
\right]^2
={}&
\left[
\frac{3D_0}{2+\alpha\beta}
\right]^2
\\
&\times
\left[
\beta^2\frac{R-x}{L}
+
\alpha^2\frac{x-\bar r}{L}
\right].
\end{split}
\label{eq:app_square}
\end{equation}

For $0<\alpha<1$, taking the positive branch, which matches
the endpoint values, we obtain
\begin{equation}
D_{\rm opt}(x)
=
\frac{3D_0}{2+\alpha\beta}
\Bigg[
\alpha\beta
+
\sqrt{\beta^2\frac{R-x}{L}
+
\alpha^2\frac{x-\bar r}{L}
}
\Bigg].
\label{eq:app_optimal_profile}
\end{equation}
This reproduces Eq.~\eqref{eq:optimal_profile_1d} in the main text.
The It\^o and H\"anggi--Klimontovich profiles follow by taking
the limits $\alpha\to0$ and $\alpha\to1$, respectively.

\section{Numerical optimization: Constrained gradient descent}
\label{ap:gradient}
For the numerical optimization, we use constrained gradient descent,
also known as projected gradient descent~\cite{bertsekas1995nonlinear}.
We first discretize the expression for the MFPT in Eq.~\eqref{eq:mfpt}.
We write $D(r)=u(r)D_0$ and divide the interval $[\bar r,R]$
into $N$ subintervals, with endpoints
$\bar r=\ell_0<\ell_1<\dots<\ell_N=R$.
We choose these points so that $r_0=\ell_m$.
In each subinterval $[\ell_i,\ell_{i+1})$, we approximate $u(r)$
by a constant $u_i$, for $i=0,\dots,N-1$.
The discretized MFPT becomes
\begin{equation}
    T = \frac{1}{D_0}\sum_{i=m}^{N-1}
    \left[
        \frac{C_i}{u_i}
        + G_i u_i^{-\alpha}
        \sum_{j=0}^{i-1}V_j u_j^{\alpha-1}
    \right],
\end{equation}
where $V_j=(\ell_{j+1}^d-\ell_j^d)/d$,
\begin{equation*}
    G_i=
    \begin{cases}
        \displaystyle
        \ln\left(\frac{\ell_{i+1}}{\ell_i}\right),
        & d=2, \\[10pt]
        \displaystyle
        \frac{\ell_{i+1}^{2-d}-\ell_i^{2-d}}{2-d},
        & d\neq 2,
    \end{cases}
\end{equation*}
and $C_i=\left[(\ell_{i+1}^2-\ell_i^2)/2-\ell_i^d G_i\right]/d$.

We initialize with the homogeneous profile,
$u_i^{(0)}=1$ for $i=0,\dots,N-1$, corresponding to $D_i=D_0$.
At each iteration, we move in the direction opposite to
the gradient of $T$,
\begin{equation}
    v_i^{(k)}
    = u_i^{(k)} - h_k
    \frac{\partial T}{\partial u_i}
    \left(u_0^{(k)},\dots,u_{N-1}^{(k)}\right),
\end{equation}
where $h_k>0$ is the step size. Then, we seek the 
profile closest to $(v_0^{(k)},\dots,v_{N-1}^{(k)})$ that satisfies the constraints.
To do this, we minimize the sum of squared differences
\begin{equation}
    (u_0^{(k+1)},\dots,u_{N-1}^{(k+1)})
    = \underset{u_0,\dots,u_{N-1}}{\arg\min}\,
    \frac{1}{2}\sum_{i=0}^{N-1}
    \left(u_i-v_i^{(k)}\right)^2,
\end{equation}
subject to $\sum_{i=0}^{N-1}w_i u_i=1$ and $u_i\geq\epsilon$,
where
\begin{equation}
    w_i=\frac{\ell_{i+1}^d-\ell_i^d}{R^d-\bar r^d}.
\end{equation}
The first constraint keeps the volume-averaged diffusivity
equal to $D_0$. The second keeps the diffusivity positive,
where $\epsilon$ is a small positive number less than one.
If the lower bound is not active, we can solve the minimization
using a Lagrange multiplier. We define
\begin{equation}
    \mathcal{L}_u
    = \frac{1}{2}\sum_{i=0}^{N-1}
    \left(u_i-v_i^{(k)}\right)^2
    + \lambda\left(\sum_{i=0}^{N-1}w_i u_i-1\right).
\end{equation}
The conditions $\partial_{u_i}\mathcal{L}_u=0$ and
$\partial_{\lambda}\mathcal{L}_u=0$ give
\begin{equation}
    u_i^{(k+1)}
    = v_i^{(k)}-\lambda w_i,
    \qquad
    \lambda
    = \frac{\sum_{i=0}^{N-1}w_i v_i^{(k)}-1}
    {\sum_{i=0}^{N-1}w_i^2}.
\end{equation}
This solution is valid if $u_i^{(k+1)}\geq\epsilon$ for all $i$.
Otherwise, we set the components below $\epsilon$ equal to
$\epsilon$ and use $u_i^{(k+1)}=v_i^{(k)}-\lambda w_i$
for the remaining components. We recalculate $\lambda$ so that
$\sum_{i=0}^{N-1}w_i u_i^{(k+1)}=1$ and repeat this adjustment
until all components satisfy $u_i^{(k+1)}\geq\epsilon$.
The gradient descent iterations continue until
the convergence criterion is reached.
\section{Langevin simulation}
\label{app:langevin}
For the one-dimensional Langevin simulations with the optimal
diffusivity profile $D_\star(x)$, trajectories were generated using
the Euler's method of the corresponding It\^o equation,
\begin{equation}
    x_{n+1} = x_n + \alpha D_\star'(x_n)\Delta t
    + \sqrt{2D_\star(x_n)\Delta t}\,\xi_n,
\end{equation}
where $\xi_n$ are independent standard Gaussian random variables.
Each particle started at $x=0$. Reflection at this boundary was
implemented by replacing negative positions with their absolute
values, and the trajectory was terminated when $x_{n+1}\geq L$,
recording the first-passage time as $(n+1)\Delta t$.
We used $\Delta t=10^{-4}$, $L=5$, and $D_0=1$, with $10^5$
independent trajectories for each value of $\alpha$.
The mean first-passage time was estimated by averaging these times,
and the efficiency was computed as
$\eta=1-\langle T\rangle/[L^2/(2D_0)]$, for comparison with
$\eta=(2\alpha-1)^2/9$.
Error bars represent the standard error of the sample mean divided
by $L^2/(2D_0)$.

\bibliography{references}

\end{document}